\documentclass[intlimits,twoside,a4paper]{article}
\usepackage[utf8]{inputenc}
\usepackage{graphicx}
\usepackage{braket}
\usepackage{upgreek}
\usepackage{cmpj3}
\newcommand{\vvec}[1]{\mathbf{#1}}

\issue{2023}{26}{3}{33302}
\doinumber{10.5488/CMP.26.33302}

\title[Modelling thermoresponsive polymer brush by mesoscale computer simulations]{Modelling thermoresponsive polymer brush by mesoscale computer simulations}
\author[D. Yaremchuk, O. Kalyuzhnyi, J. Ilnytskyi]{
D. Yaremchuk\orcid{0000-0003-2888-5878},
O. Kalyuzhnyi\orcid{0000-0002-5691-3058},
J. Ilnytskyi\orcid{0000-0002-1868-5648}\thanks{Corresponding author: \email{iln@icmp.lviv.ua}.}
}

\address{
Institute for Condensed Matter Physics of the National Academy of Sciences of Ukraine,\\
1 Svientsitskii Str., 79011, Lviv, Ukraine
}


\date{Received April 17, 2023}

\begin{document}
\maketitle
\begin{abstract}

We consider a functional surface comprising thermoresponsive polymer chains, the material that has found numerous technological and biomedical applications. However, to achieve the required time and length scales for computer modelling of such applications, one is compelled to use coarse-grained mesoscopic modelling approaches. The model used here is based on the previous work [Soto-Figueroa et al., Soft Matter, {\textbf 8}, 1871 (2012)], and it mimics the principal feature of the poly(N-iso-propylacrylamide) (PNIPAM), namely, the rapid change of its hydrophilicity at the lower critical solution temperature (LCST). For the case of an isolated chain, we discuss scaling properties of the radius of gyration, end-to-end distance, various distribution functions, and the density profile of monomers below and above the LCST. For the case of the model thermoresposive brush, we search for the optimum grafting density at which the change in the brush height, upon crossing the LCST, reaches its maximum value. The interpretation of the thermoresponse, in terms of the Alexander-de Gennes blobs and the level of solvation of polymer chains in a brush, is provided.
\printkeywords
\end{abstract}

\section{Introduction}\label{sec1}

Stimuli-responsive polymers have a wide range of applications in biotechnology and medicine~\cite{Cabane2012}. Thermoresponsive polymers form a sub-class of these materials, and the PNIPAM is their prominent representative. Applications of this polymer are based on a drastic change of its solubility in water that occurs around the LCST. Below LCST, PNIPAM is hydrophilic and it swells in water, whereas above LCST, PNIPAM chains collapse~\cite{Graziano2000}. The LCST for this polymer is close to $32^{\circ}$C, which is important from the point of view of the physiological applications~\cite{Nagase2018}.

Stability of a swollen state of PNIPAM below LCST is attributed to the formation of water cages and hydrogen bond bridges with water molecules~\cite{Cho2003,Ono2007}. Upon the temperature increase, the number of PNIPAM–water hydrogen bonds decrease, whereas the number of the polymer–polymer hydrogen bonds increase, accompanied by desolvation of the hydrophobic groups, as suggested by spectroscopy studies~\cite{Maeda2000,Sun2008,Ahmed2009,Lai2010}. The desolvation of the polymer above the LCST  manifests itself in an essential decrease in the hydration number~\cite{Ono2006,Philipp2014}. 

This microscopic picture is confirmed in a number of computer simulation studies of a single PNIPAM chain using a fully atomistic modelling approach~\cite{Tavagnacco2018,Dalgicdir2019,Tavagnacco2021}. It was shown that a network of hydrogen bonds is formed between the PNIPAM segments and the water molecules below LCST resulting in hydration of the PNIPAM chains and their coiled conformations. Above LCST, such a network breaks in the proximity of hydrophobic isopropyl groups, and the PNIPAM chains collapse into compact globules~\cite{Tavagnacco2018}. The transition, that occurs at LCST, can therefore be classified as the conformational coil-to-globule transition. Dalgicdir et al.~\cite{Dalgicdir2019} tested the OPLS/AA force field for a single PNIPAM 40-mer in water. They found that the experimental coil-to-globule collapse enthalpy of PNIPAM in water can be reproduced by simulations forcing the amide-water electrostatic interactions stronger than in the original OPLS model. This indicates high sensitivity of the coil-to-globule transition to the fine details of the force fields. Besides the obvious role of the temperature, the pressure also plays an important role in this transition. This effect was addressed by Tavagnacco et al.~\cite{Tavagnacco2021}. In particular, a coil-to-globule transition was found to occur up to large pressures, and a new high pressure globular state was found. It is characterized by a more structured hydration shell that is closer to PNIPAM hydrophobic domains, as compared to the globular state at atmospheric pressure.

A single PNIPAM chain, grafted inside a cylindrical pore that resembles a carbon nanotube, was considered by Alaghemandi et al.~\cite{Alaghemandi2013}. To suppress the collapse of PNIPAM on a substrate, they reduced the AMBER force field interaction strengths by a factor of ten. For this case, the end-to-end distances, radii of gyration, density profiles, the number of hydrogen bonds, and radial distribution functions for a PNIPAM chain were evaluated, in order to characterize the coil-to-globule transition on a level of the monomers rearrangement. The cases of the PNIPAM containing a surfactant molecule~\cite{Abbott2015} and PNIPAM based hydrogels~\cite{Walter2010,Zanatta2020} have also been studied. In particular, Walter et al.~\cite{Walter2010} reported a fair agreement between the experimental and computer simulation studies of the coil-to-globule transition in the PNIPAM hydrogel in water. 

Atomic resolution computer simulations provide invaluable information on the chemistry detailed properties of the PNIPAM chains and underlying mechanisms of their conformational changes, as well as their hydration and  specific features of dehydration. However, most applications of the PNIPAM based systems occur at a relatively large length and time scales, be these either micellar systems for temperature controlled drug delivery or polymer brushes for smart commanding surfaces. Atomistic simulations at this scale in most cases prove to be too costly, which calls for more economical coarse-grained modelling approaches. These treat each characteristic group of atoms as a single particle, and their interaction potentials represent total interaction of their constituent atoms with their surroundings. This enables a drastic reduction of the degrees of freedom, but should be appropriately parametrized to retain the principal features of the initial PNIPAM chain. General concepts of a rigorous coarse-graining of polymers is covered in a number of reviews, e.g., in the recent one by Dhamankar and Webb~\cite{Dhamankar2021}.

Mid-level coarse-graining of the PNIPAM polymer involves treating its repeating unit as a trimer. It includes the hydrocarbon group of the main chain and a hydrophilic amide group and a hydrophobic isopropyl group of a side chain (see, e.g., references~\cite{Abbott2015a,PerezRamirez2020}). Hence, each such group contains 2--3 heavy atoms. Abbott et al.~\cite{Abbott2015a} followed the coarse-graining recipe by Shinoda et al.~\cite{Shinoda2007}, when the nonbonded interactions were fit to experimental thermodynamic data. It was necessary to derive different potentials at both temperatures, $280\,\mathrm{K}$ and $330\,\mathrm{K}$, to capture a correct temperature-dependence of the interactions. The coil-to-globule transition was observed for the longer, $N=18$ and $N=30$, oligomers only. Botan et al.~\cite{Botan2017} developed an implicit solvent coarse-grained model coupled to a lattice Boltzmann hydrodynamics, but accounting for these interactions was insufficient to capture the coil-to-globule transition. P\'{e}rez-Ram\'{i}rez and Odriozola~\cite{PerezRamirez2020} developed a similar level coarse-grained model based on the Martini force field. The amide moieties include an electric dipole, and polymer chains of different sizes accurately reproduce  the thermal behaviour without including temperature-dependent parameters. Bejagam et al.~\cite{Bejagam2018} applied a data-driven machine-learning method combined with a coarse-grained modelling to study the conformations of PNIPAM in solutions. The temperature independent model being developed accurately predicts the LCST, at the same time retaining the tacticity in the presence of an explicit water model. The nonmetric multidimensional scaling method shows the presence of multiple metastable states of PNIPAM during its coil-to-globule transition above the LCST. All in all, the coarse-grained modelling of this level has its strong points in drastic reduction of the degrees of freedom retaining the molecular topology of PNIPAM polymer and relevant details of its chemical moieties, and, therefore, may be seen as an optimum compromise in modelling PNIPAM as for now. 

It is, however, tempting, to go one step further and to examine the capabilities of yet more coarse-grained approach, like the one used in references~\cite{Ilnytskyi2011, Kuroki2019}. In this case, PNIPAM is represented as a linear chain. The logic for employing this approach is at least twofold. Firstly, the experimentally observed effect may occur on a rather large length scale and its purely physical, rather than chemical, interpretation might be sufficient. This is the case, for instance, for the setup considered in a recent paper by Minko group~\cite{Kim2021}. Here, the separation of colloid particles from a solution is achieved by a local oscillating repulsive force generated at the microstructured stimuli-responsive polymer interface, in order to switch between adsorption and mechanical-force-facilitated desorption of the particles. The polymer interface is in the form of a patterned mixed brush, with one of the components being made of the PNIPAM chains of considerable length. Below LCST, the PNIPAM areas of a brush swell and rise up, breaking the proteins out free. Considering that the swelling of a small PNIPAM region of such brush can be described by the mid-level coarse-grained models discussed in a previous paragraph, the role of the spatial pattern in the effectiveness of such smart surface clearly requires large system sizes inaccessible to such models. Here, such higher level of coarse-graining, as introduced above, would be of a great aid.

Secondly, there is already some work done on modelling the polymers that involve the PNIPAM fragments~\cite{SotoFigueroa2012,BautistaReyes2016,RodrguezHidalgo2018,Lang2018}. Justification of the effective interaction parameters between the PNIPAM monomers and water, including the formation of hydrogen bonds, was discussed by Vicente with colleagues~\cite{BautistaReyes2016}. They addressed both theoretical calculation of Flory–Huggins parameter $\chi$ and the Monte Carlo simulations approach involving the evaluation of the coordination numbers and binding energies between interacting particles. As a result, the temperature dependence of the parameter $\chi$ for the PNIPAM chain was obtained.

The aim of this study is to examine whether or not this type of modelling faithfully reproduces the main features of the PNIPAM below as well as above the LCST. Both the cases of an isolated PNIPAM chain and the ensemble of the PNIPAM chains, arranged as a polymer brush, are considered. For the former, we examine the scaling behaviour of a number of its properties with respect to the chain length. For the latter, we focus on a search for an optimum grafting density of a brush at which the thermoswitching effect has the maximum magnitude, and also discuss microscopic mechanisms behind the thermoswitching behaviour.

This research work is dedicated to Taras Bryk, leading expert in the dynamic properties of liquids, liquid metals and alloys, a specialist in the ab initio and classical molecular dynamics simulations, Director of the Institute for Condensed Matter Physics, on the occasion of his 60th birthday. We wish him scientific prolificity, new ideas and achievements for the years to come.

The outline of the study is as follows. In section~\ref{sec2} we provide details of the model and of the simulation technique, the properties of an isolated PNIPAM chain are discussed in section~\ref{sec3}, whereas section~\ref{sec4} contains the results for the case of the PNIPAM brush at different grafting densities. The study is summarised by conclusions.

\section{Modelling details}\label{sec2}

We use the dissipative particle dynamics (DPD) simulation approach, which belongs to the class of mesoscopic simulation techniques. In doing this we closely follow  the seminal work by Groot and Warren~\cite{Groot1997}. In this approach, the monomers of a polymer chain, as well as the solvent, are all represented as soft spheres of equal size that interact via pairwise forces. The diameter of the soft bead defines a native length-scale in a model, and the energy unit is chosen equal to $k_{\rm B}T=1$, where $k_{\rm B}$ is the Boltzmann constant and $T$ is the temperature, time unit is $t^*=1$. The monomers are bonded via harmonic springs which results in the bonding force:
\begin{equation}\label{FB}
  \vvec{F}^B_{ij} = -k\vvec{x}_{ij},
\end{equation}
where $\vvec{x}_{ij}=\vvec{x}_i-\vvec{x}_j$, $\vvec{x}_i$ and $\vvec{x}_j$ are the positions of the $i$-th and $j$-th bead, and $k$ is the associated energy. The total non-bonded force $ \vvec{F}_{ij}$ acting on $i$-th bead from its $j$-th counterpart can be expressed as a sum of three contributions:
\begin{equation}
  \vvec{F}_{ij} = \vvec{F}^{\mathrm{C}}_{ij} + \vvec{F}^{\mathrm{D}}_{ij}
  + \vvec{F}^{\mathrm{R}}_{ij},
\end{equation}
where $\vvec{F}^{\mathrm{C}}_{ij}$ is the conservative force caused by the repulsion of the $i$-th and $j$-th beads, $\vvec{F} ^{\mathrm{D}}_{ij}$ is the dissipative force responsible for
friction between beads (similar to the Stokes force), and the random force $\vvec{F}^{\mathrm{R}}_{ij}$ which, together with the dissipative force, maintains a constant temperature of the system. Expressions for these terms are pretty standard and are given below:
\begin{equation}\label{FC}
  \vvec{F}^{\mathrm{C}}_{ij} =
     \left\{
     \begin{array}{ll}
        a(1-x_{ij})\displaystyle\frac{\vvec{x}_{ij}}{x_{ij}}, & x_{ij}<1,\\
        0,                       & x_{ij}\geqslant 1,
     \end{array}
     \right.
\end{equation}
\begin{equation}\label{FD}
  \vvec{F}^{\mathrm{D}}_{ij} = -\gamma
  w^{\mathrm{D}}(x_{ij})(\vvec{x}_{ij}\cdot\vvec{v}_{ij})\frac{\vvec{x}_{ij}}{x^2_{ij}},
\end{equation}
\begin{equation}\label{FR}
  \vvec{F}^{\mathrm{R}}_{ij} = \sigma
  w^{\mathrm{R}}(x_{ij})\theta_{ij}\Delta t^{-1/2}\frac{\vvec{x}_{ij}}{x_{ij}},
\end{equation}
where $x_{ij}=|\vvec{x}_{ij}|$, $\vvec{v_{ij}=\vvec{v_i}}-\vvec{v_j}$, $\vvec{v_i}$ is the velocity of the $i$-th monomer, and $a$ is the amplitude for the conservative repulsive force. The dissipative force has an amplitude $\gamma$ and its decrease with the particle separation is given by the function $w^{\mathrm{D}}(x_{ij})$. Similarly, the amplitude for the random force is $\sigma$ and its decrease is provided by $w^{\mathrm{R}}(x_{ij})$. $\theta_{ij}$ is a Gaussian random variable. As shown in~\cite{Espanol1995},  to satisfy the requirement of a detailed balance, the relation $\sigma^2 = 2\gamma$ and $w^{\mathrm{D}}(x_{ij})=\large[w^{\mathrm{R}}(x_{ij})\large]^2$ should hold. Here, we use a quadratic decaying form for the weight function:
\begin{equation}\label{WD}
  w^{\mathrm{D}}(x_{ij})=\left[w^{\mathrm{R}}(x_{ij})\right]^2 =
     \left\{
     \begin{array}{ll}
        (1-x_{ij})^2, & x_{ij}<1,\\
        0,          & x_{ij}\geqslant 1.
     \end{array}
     \right.
\end{equation}

\begin{figure}[htb]
	\centering
	\includegraphics[width=0.45\linewidth]{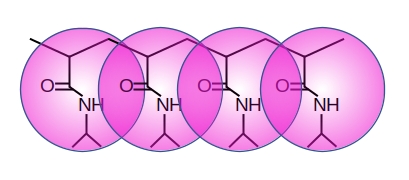}   
	\caption{\label{fig:PNIPAM}(Colour online) Schematic representation of the PNIPAM chain fragment modelled as a linear chain of coarse-grained beads.}
\end{figure}

Mesoscopic modelling of the PNIPAM polymer follows closely the work by Soto-Figueroa et al.~\cite{SotoFigueroa2012}, where the PNIPAM-containing block-copolymers were considered. Namely, each repeating unit of the PNIPAM is represented via a single soft spherical bead, as shown in figure~\ref{fig:PNIPAM}. Each polymer chain consists of $N$ such monomers that are bonded pairwise via the force (\ref{FB}) with the spring constant $k=4$.

In general, repulsive strength of the conservative force, $a$, in equation~(\ref{FC}) depends on the temperature and density. In their practical implementation of the DPD approach, Groot and Warren considered the fixed bulk number density of beads equal to $3$, and parameter $a$ was found from matching the inverse compressibility of a model to that of water at normal conditions~\cite{Groot1997}. This resulted in the value of $a=25$. We denote polymer beads by a subscript ``$p$'', and those representing water, by ``$w$''. It is assumed that for the polymer-polymer and water-water interaction, one has $a_{pp}=a_{ww}=25$. Mixed, polymer-water interaction strength, $a_{pw}$, depends on the polymer level of hydrophilicity. In the case of PNIPAM, one expects $a_{pw}\approx 25$ below LCST and $a_{pw}>25$ above it. Hence, the strength of the polymer-water hydrogen bonding does enter this model implicitly, via the value of the $a_{pw}$. The temperature dependence of $a_{pw}$ was estimated from that for the Flory-Huggins mixing parameter $\chi$, see, figure~3 in~\cite{SotoFigueroa2012}, using the relation between the two~\cite{Groot1997}. To simplify the analysis, in this study we concentrate on two temperatures only: one below, and another one above the LCST. Namely, at $T=298\,\textrm{K}$ (below LCST), the estimate for $a_{pw}$ is $a_{pw}\approx 25.6$, whereas at $T=310\,\textrm{K}$ (above LCST), one has $a_{pw}\approx 38$.

The simulation setup is as follows. The simulation box dimensions are $L_x$, $L_y$ and $L_z$, which are  specified in each case being considered. The periodic boundary conditions are applied along both OX and OY axes, whereas two walls representing a substrate are imposed at $z=0$ and $z=L_z$. Monomer-surface interaction plays a big role in specifying the structure of a brush at various densities~\cite{Descas2008}. Here, we restricted our attention to a singular case of repulsive impenetrable walls. The repulsion from the wall is of the same soft nature as between the beads and is realized using the expression (\ref{FC}). Here, $\vvec{x}_{ij}$ denotes the vector that runs from the center of $i$-th bead towards the nearest wall and is normal to it. The amplitude of the bead-wall repulsion is fixed at $a_{pw}=25$. This type of repulsion does not ensure the impenetrability of the walls. Therefore, the elastic reflection is imposed when the bead crosses the either of the walls.  

In general, we consider a certain number $M$ of chains, and one of the end beads of each chain is grafted to a specified fixed point of the wall located at the $z=0$ plane. Grafting is performed by using the harmonic force (\ref{FB}) with $k=4$. If grafting points are chosen randomly, then one ought to average over many such arrangements and, additionally, the results are affected by local inhomogeneities of the surface density of grafting points. To reduce these effects, we opted to arrange the grafting points on a regular grid and perform a single simulation run for each $M$. Specifically, the grid of $K\times K$ grafting points is made in the form of a square lattice, where $K$ is the integer value of $\sqrt{M}$. The rest $M-K^2$ grafting points are distributed randomly across the wall. The grafting density is measured as the surface number density of grafts, $\rho_g=M/(L_xL_y)$. The chains are not allowed to slide on the surface. The rest of the simulation box is filled-in with solvent beads to achieve a required total bulk density of beads equal to $3$~\cite{Groot1997}. The time step was chosen equal to $\Delta t=0.04$ in the DPD units. For the case of an isolated chain, $M=1$, considered in section~\ref{sec3}, $8\cdot 10^6$ DPD steps are performed, where the first $10^6$ of them are discarded to allow conformational relaxation of the chain. For the case of a brush, $M>1$, covered in section~\ref{sec4}, $10^6$ DPD steps were used, where first $3\cdot 10^5$ of them were discarded for relaxation purpose.

\section{Isolated grafted thermoresponsive chain}\label{sec3}

Scaling properties of polymer chains, initiated in the seminal works by Flory, des Cloizeaux, de Gennes~\cite{Flory,desCloizeaux1982,deGennes1980}, are considered to be a well studied topic now. These studies brought the concept of universality into the field of polymer physics, mirroring a self-similarity of polymeric conformations. The universality features hold especially well at a mesoscopic level, where the fine chemical details are smeared out because of coarse-graining. Some examples can be found in our previous studies~\cite{Ilnytskyi2007,Kalyuzhnyi2016,Kalyuzhnyi2019,Haydukivska2021,Haydukivska2022}. As was discussed in section~\ref{sec2}, the PNIPAM polymer can be mapped onto a linear chain with the polymer-water interactions specified by the magnitude of $a_{pw}$ in the expression of a soft repulsive force (\ref{FC}). Below LCST, the value of $a_{pw}\approx 25.6$ is found~\cite{SotoFigueroa2012}, which is pretty close to the case of athermal solvent, $a_{pw}=a_{pp}=a_{ww}=25$. Such case was extensively studied before, see, e.g.~\cite{Ilnytskyi2007,Kalyuzhnyi2016} and references therein. The aim of this study is not to merely repeat these findings but to perform a detailed comparison between the properties of an isolated chain at $T=298\,\mathrm{K}$ (below LCST) and at $310\,\mathrm{K}$ (above it). Such comparison, that is seen as the first step in studying the thermoresponsive brush to be performed in section~\ref{sec4}, was not done in detail before.

We consider an isolated chain that is grafted to the plane $z=0$ and surrounded by a solvent that represents water. The details of modelling are provided in section~\ref{sec2}. The set of chain lengths, $N=10$, $14$, $20$, $28$, $40$, and $56$, are considered in the three dimensional space. The simulation box dimensions in each case is chosen: (i) to be large enough to accommodate the most probable conformations of a chain, and (ii) not to be oversized to save on a computer time. Some recipes were suggested earlier~\cite{Kalyuzhnyi2016}, and in the current study we found a good compromise by the following choice: $L_x=L_y=L_z=b_0N^{0.8}$, where $b_0\approx 0.9$~\cite{Ilnytskyi2007} is an equilibrium bond length between the monomers for the chosen values of $k$ and $a$ in equations~(\ref{FB}) and (\ref{FC}) and a bulk number density of $3$. In parallel, we perform similar simulations for a free chain of the same length at the same solvent conditions. The periodic boundary conditions are applied along all three spatial axes in this case. Simulation runs are of the duration of $8\cdot10^6$ DPD steps for each chain length $N$, and the first $10^6$ steps are discarded from the analysis to allow equilibration of the system.\vspace{0.5cm}

Let us consider the size properties of a polymer chain. Most generally, the spatial distribution of its mass is given by the gyration tensor, $\vvec{G}$, with its components defined as~\cite{Stockmayer}
\begin{equation}\label{eq:gyr_tens}
G_{\alpha\beta} = \frac{1}{N}\sum_{i=1}^{N} (r_{i,\alpha}-R_{\alpha}) (r_{i,\beta}-R_{\beta}).
\end{equation}
Here, $r_{i,\alpha}$ and $R_{\alpha}$ denote Cartesian coordinates of $i$-th monomer and these for the centre of mass of a chain, respectively, and $\alpha=1,2,3$. 

\begin{figure}[htb]
  \centering
\hspace{-2em}
\includegraphics[width=0.5\linewidth]{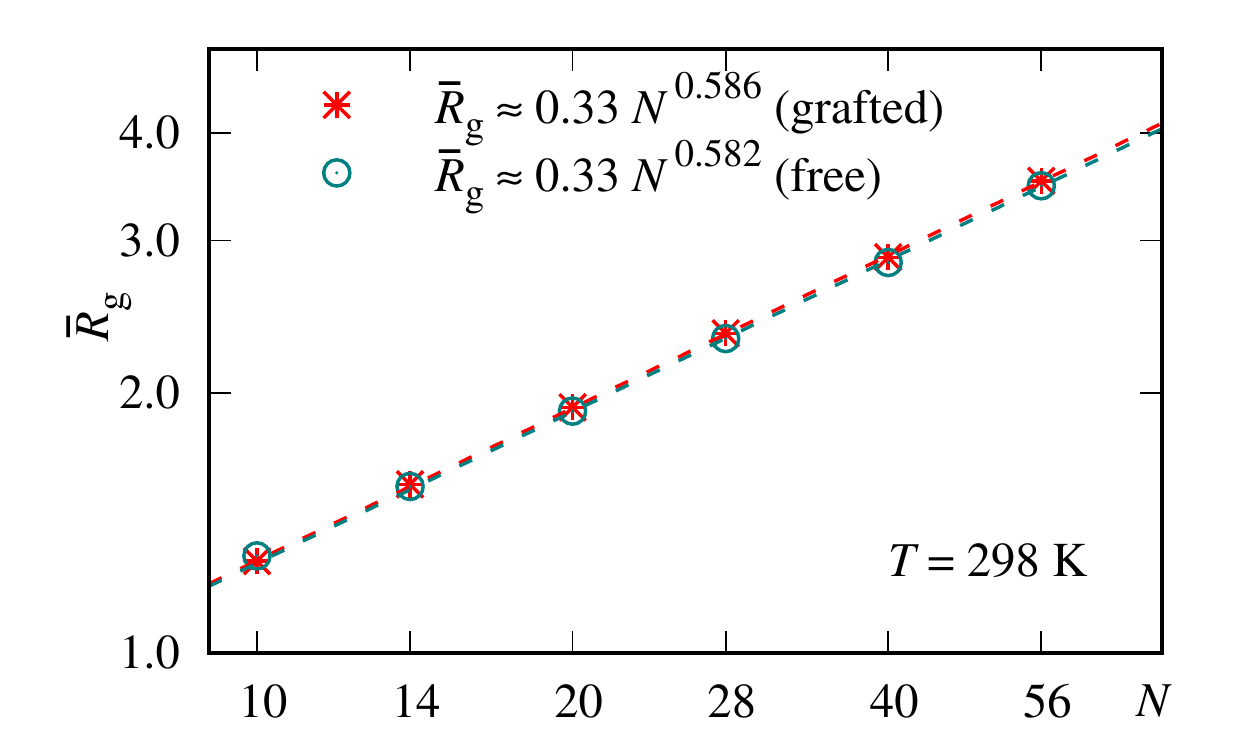}\hspace{-2em}
\includegraphics[width=0.5\linewidth]{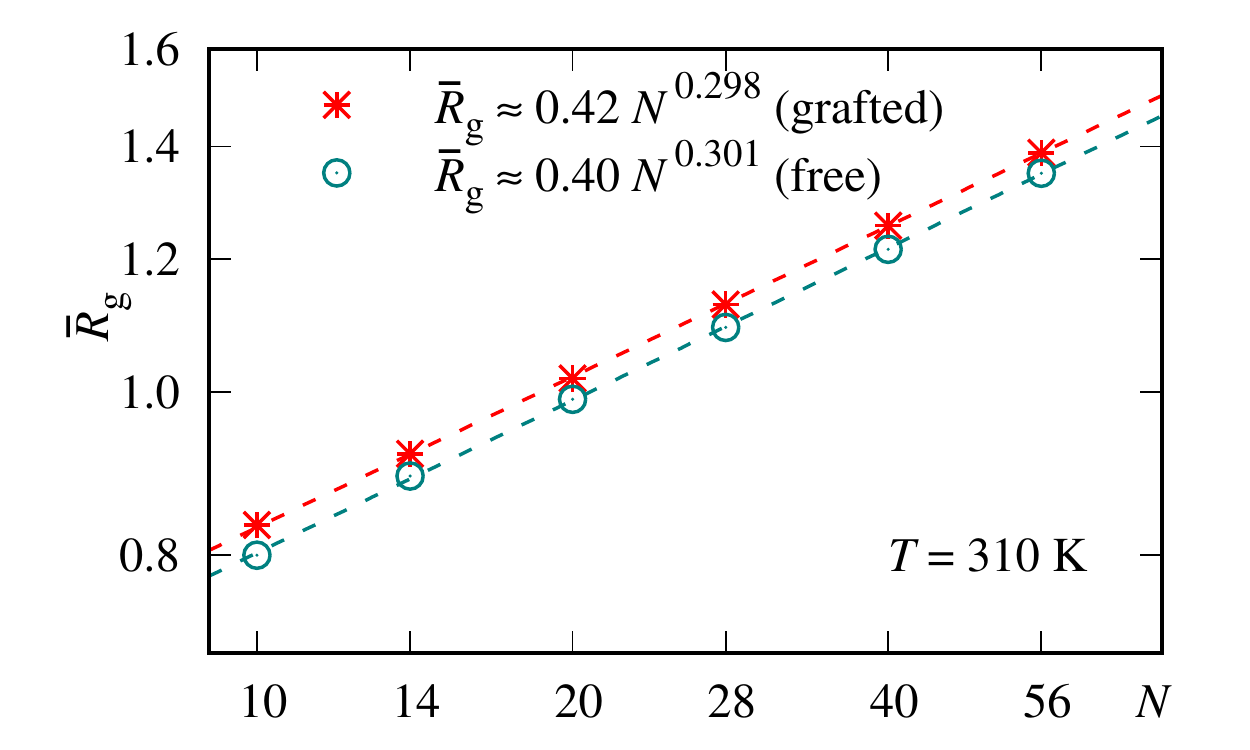}
\caption{\label{fig:Rg_scaling_single}(Colour online) Scaling of the average radius of gyration $R_g$ of a single model PNIPAM chain with the number of monomers, $N$, at two temperatures, $T=298\,\mathrm{K}$ and $310\,\mathrm{K}$. DPD data are displayed via symbols. Both cases of a free and grafted chains are shown and the results for their fits to the power law~(\ref{eq:Rg_scaling}) are provided in the plots.}
\end{figure} 
The eigenvalues $\lambda_{\alpha}$ of the gyration tensor are sorted as follows $\lambda_1>\lambda_2>\lambda_3$. These allow one to evaluate the instantaneous squared radius of gyration of a chain, $R_g^2=\sum_{\alpha=1}^{3}\lambda_{\alpha}$. The average value for the radius of gyration is defined as  follows:
\begin{equation}\label{eq:Rg_scaling}
\bar{R}_g = \langle R_g^2 \rangle^{1/2} \sim l\, N^{\nu},
\end{equation}
where the averaging is performed over the time trajectory and we already used the scaling law with respect to the number of monomers, $N$,~\cite{deGennes1980,desCloizeaux1982}. Here, $l$ is a pre-factor, and $\nu$ is the scaling exponent which depends on the solvent quality. In particular, $\nu\approx 0.5882$~\cite{Guida1998} for the case of good solvent, $\nu=1/2$ for the random walk (Gaussian chain) and  $\nu=1/3$ for the case of poor solvent. Grafted chain is expected to obey the power law with the same exponent $\nu$ as its free counterpart but with a different pre-factor~$l$~\cite{Binder2012}.

Simulation data obtained for $\bar{R}_g$ and their numeric fits to the from (\ref{eq:Rg_scaling}) are shown in figure~\ref{fig:Rg_scaling_single}. In each case, the data are displayed in a log-log scale to allow clear visualisation of the quality of the fits represented as straight lines. The first obvious thing is that $\bar{R}_g$ is essentially higher at $T=298\,\mathrm{K}$ than at $310\,\mathrm{K}$ (where the chain is in a state of collapse). This will have important consequences for the case of a polymer brush, as discussed later, in section~\ref{sec4}. All data shown in figure~\ref{fig:Rg_scaling_single} are found to fit rather well to the scaling law (\ref{eq:Rg_scaling}), providing a robust estimate for the critical exponent $\nu$. The estimates for $\nu\approx 0.582-0.586$ at $T=298\,\mathrm{K}$ are in a very good agreement with the known most accurate  result $\nu\approx 0.5882$~\cite{Guida1998} for the Flory exponent. At $T=310\,\mathrm{K}$, however, the value of $\nu\approx 0.300$ is close to but lower than the expected exact result, $\nu=1/3$, indicating a strong mutual penetration of monomers because of their soft nature. Grafting a chain to a  surface indeed affects the value of a pre-factor $l$ but the difference is rather minor, i.e., within the accuracy of simulations.

\begin{figure}[htb]
  \centering
\hspace{-2em}
\includegraphics[width=0.5\linewidth]{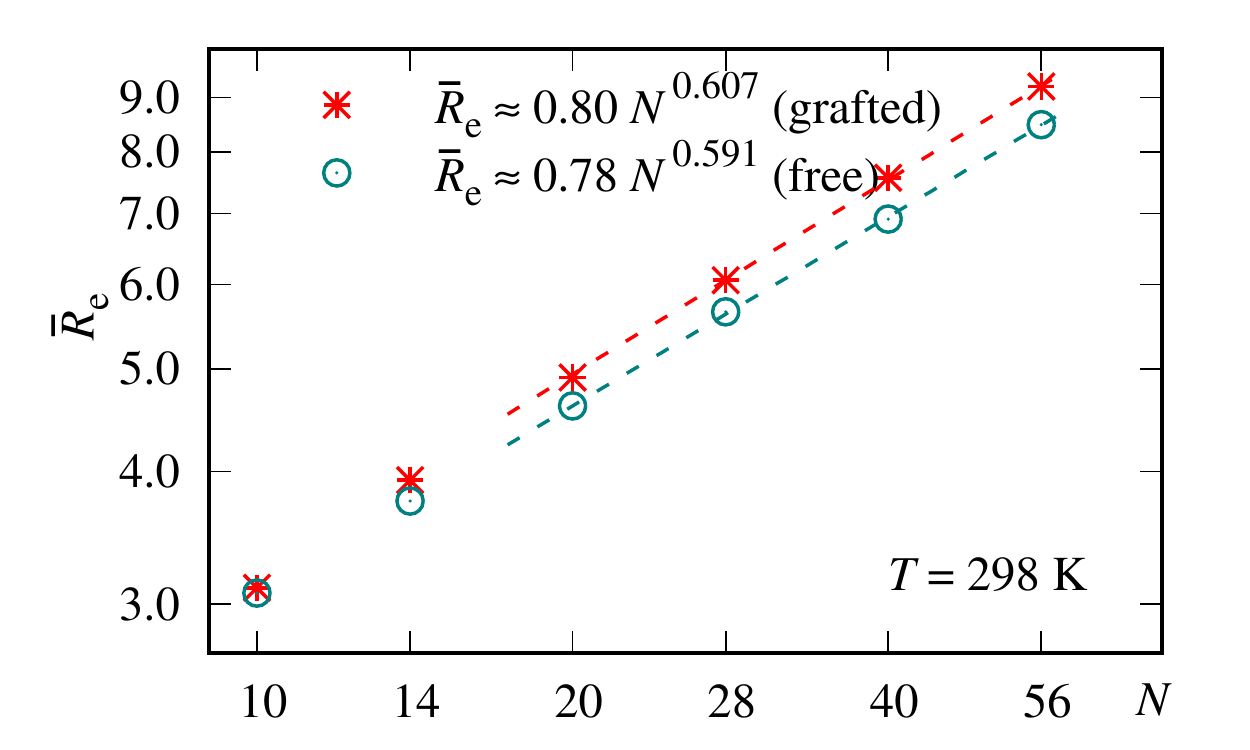}\hspace{-2em}
\includegraphics[width=0.5\linewidth]{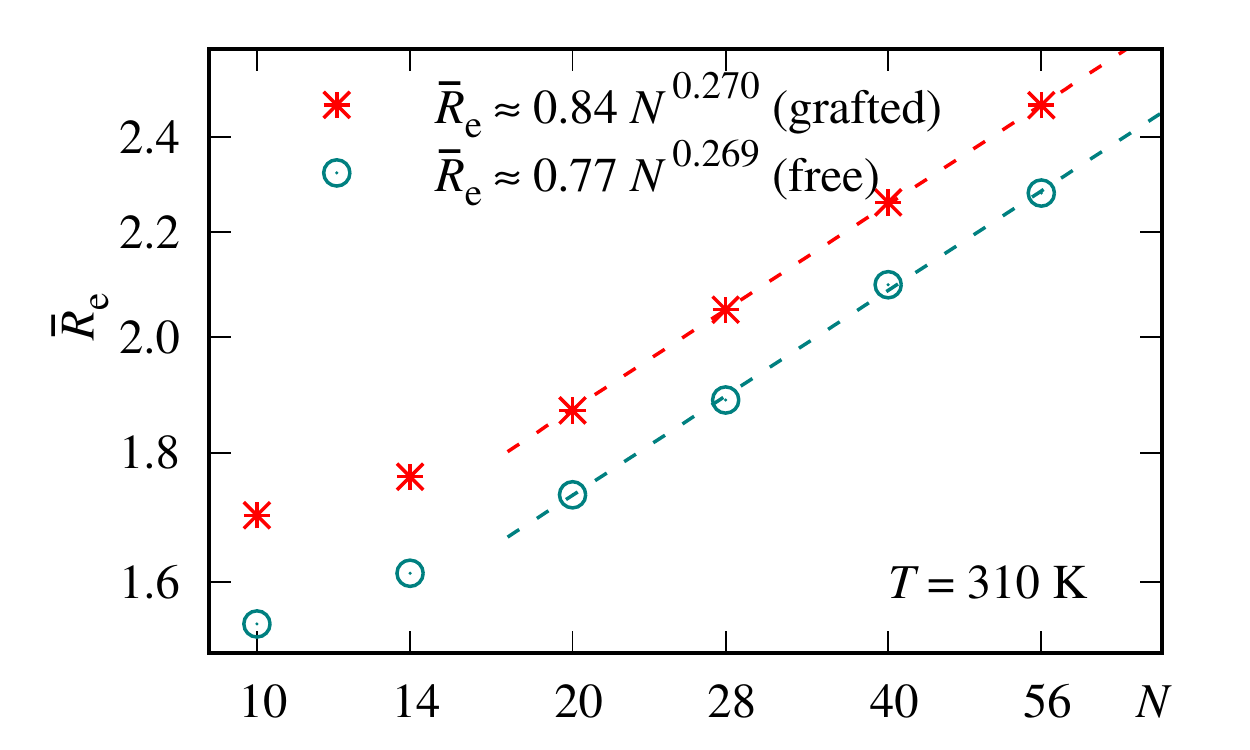}
\caption{\label{fig:Re_scaling_single}(Colour online) Scaling of the chain length $R_e$ of a single model PNIPAM chain with the number of monomers, $N$, at two temperatures, $T=298\,\mathrm{K}$ and $310\,\mathrm{K}$. DPD data are displayed via symbols. Both cases of a free and grafted chains are shown and the results for their fits to the power law (\ref{eq:Re_scaling}) are provided in the plots.}
\end{figure} 
Similar results are obtained for the average end-to-end distance of a chain defined via
\begin{equation}\label{eq:Re_scaling}
\bar{R}_e = \langle R_e^2 \rangle^{1/2} \sim l\, N^{\nu},
\end{equation}
where $R_e^2=\sum_{\alpha=1}^{3}(r_{1,\alpha}-r_{N,\alpha})^2$ and the averaging is performed over the time trajectory. Here, the pre-factor $l$ and the critical exponent $\nu$ have the same meaning as in equation~(\ref{eq:Rg_scaling}). In general, the end-to-end distance is found to be more demanding in terms of the simulation length and it displays a scaling behaviour starting from longer chain lengths, $N\geqslant 20$. The estimates for $\nu$ are acceptable for the case of $T=298\,\mathrm{K}$, whereas the value $\nu\approx 0.27$ at $T=310\,\mathrm{K}$ is even lower than that obtained from the analysis of~$\bar{R}_g$.

\begin{figure}[htb]
  \centering
\hspace{-2em}
\includegraphics[width=0.5\linewidth]{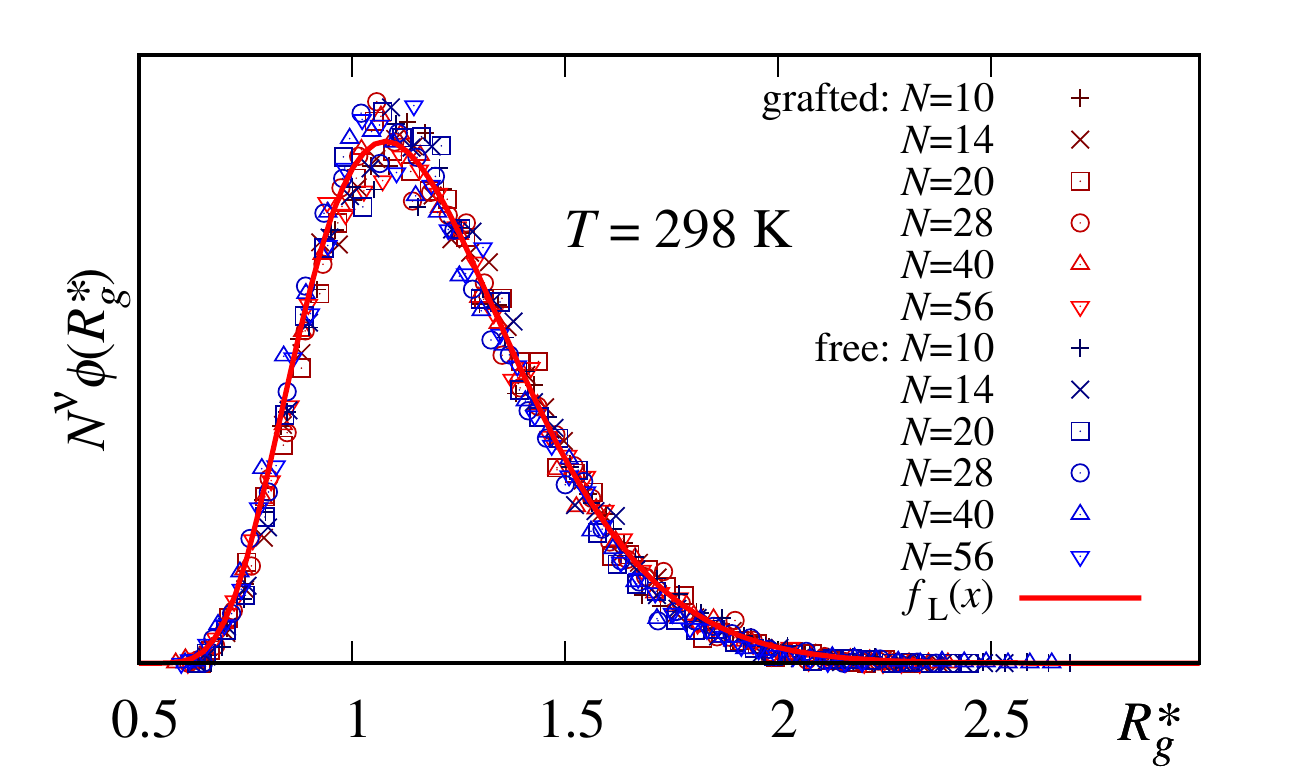}\hspace{-2em}
\includegraphics[width=0.5\linewidth]{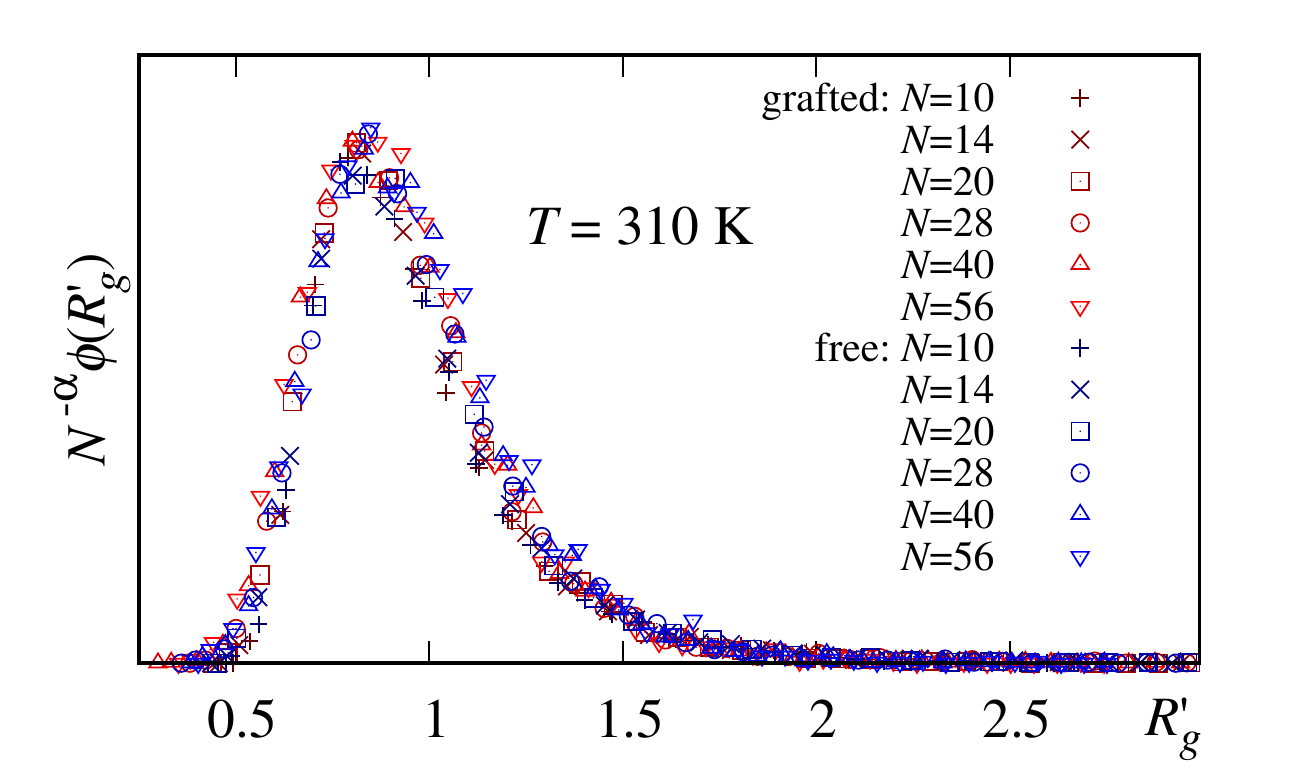}
\caption{\label{fig:Rg_distr}(Colour online) Left-hand frame: scaled distribution $N^{\nu}\phi(R^*_g)$ for the dimensionless radius of gyration $R^*_g=R_g/\bar{R}_g$ at $T=298\,\mathrm{K}$. Here, the respective scaling laws for $\bar{R}_g$, as found in the left-hand frame of figure~\ref{fig:Rg_scaling_single}, are used. Right-hand frame: scaled distributions $N^{-\alpha}\phi(R'_g)$ for dimensionless radius of gyration $R'_g$ at $T=310\,\mathrm{K}$. The latter is obtained by shifting the $R_g$: $R'_g=1+(R_g-\bar{R}_g)N^{\beta}$. The pair of unknown exponents $(\alpha,\beta)$ are found numerically as $(0.19,0.3)$ and $(0.25,0.36)$ for a grafted and free chain, respectively, and the scaling laws for $\bar{R}_g$ are used as shown in the right-hand frame of figure~\ref{fig:Rg_scaling_single}.}
\end{figure} 
We switch our attention now to the distributions of the $R_g$ values during the simulation time. For the good solvent case, $T=298\,\mathrm{K}$, such distribution can be written in terms of the dimensionless radius of gyration,  $R^*_g=R_g/\bar{R}_g$, where the scaling law for $\bar{R}_g$ is provided in figure~\ref{fig:Rg_scaling_single}. The data for different chain length $N$ obey the same master curve for the scaled distributions
\begin{equation}\label{eq:Lhuillier}
N^{\nu}\phi(R^*_g)=f_L(R^*_g)= A_L \exp\left(-\frac{1}{[R^*_g]^{\frac{3}{3\nu-1}}}-[R^*_g]^{\frac{1}{1-\nu}} \right),
\end{equation}
as shown in figure~\ref{fig:Rg_distr}. The function $f_L(x)$ is suggested by Lhuillier~\cite{Lhuillier1988}, following his considerations on the probabilities of fully collapsed and fully stretched conformations. Here, $A_L\approx 42$ is the pre-factor, the first term within the parenthesis is dominant at small $R^*_g$, whereas the second term defines the rate at which $f_L(R^*_g)$ decays at large $R^*_g$. We also would like to point out some details of our previous analysis of this distribution for the case of a free chain~\cite{Ilnytskyi2007,Kalyuzhnyi2016}.

Scaled distributions are also examined for the case of a poor solvent, $T=310\,\mathrm{K}$. We observed that the initial distributions, in terms of $R_g$, shift towards larger $R_g$ and increase moderately in their height upon the increase of the chain length, $N$. Therefore, the following expression for the shifted radius of gyration, $R_g'$, is proposed in this case (note that in the DPD simulations the length unit is assumed to be equal to unity)
\begin{equation}
R'_g = 1 + (R_g-\bar{R}_g)N^{\beta},
\end{equation}
and the distribution scales as $N^{-\alpha}\phi(R^*_g)$. The pair of unknown exponents, $(\alpha,\beta)$, is obtained by numeric fitting, yielding the values $(0.19,0.3)$ and $(0.25,0.36)$ for a grafted and free chain, respectively. The scaling laws shown in the right-hand frame of figure~\ref{fig:Rg_scaling_single}, are used for $\bar{R}_g$. The results for scaled distributions $N^{-\alpha}\phi(R^*_g)$ are presented in the right-hand frame of figure~\ref{fig:Rg_distr} and one may note a rather good match of the simulation data for different $N$. The respective master curves at both temperatures, $T=298\,\mathrm{K}$ and $310\,\mathrm{K}$, are remarkably similar in their shape, but we did not explore this similarity in any deeper terms here.

\begin{figure}[htb]
  \centering
\hspace{-2em}
\includegraphics[width=0.5\linewidth]{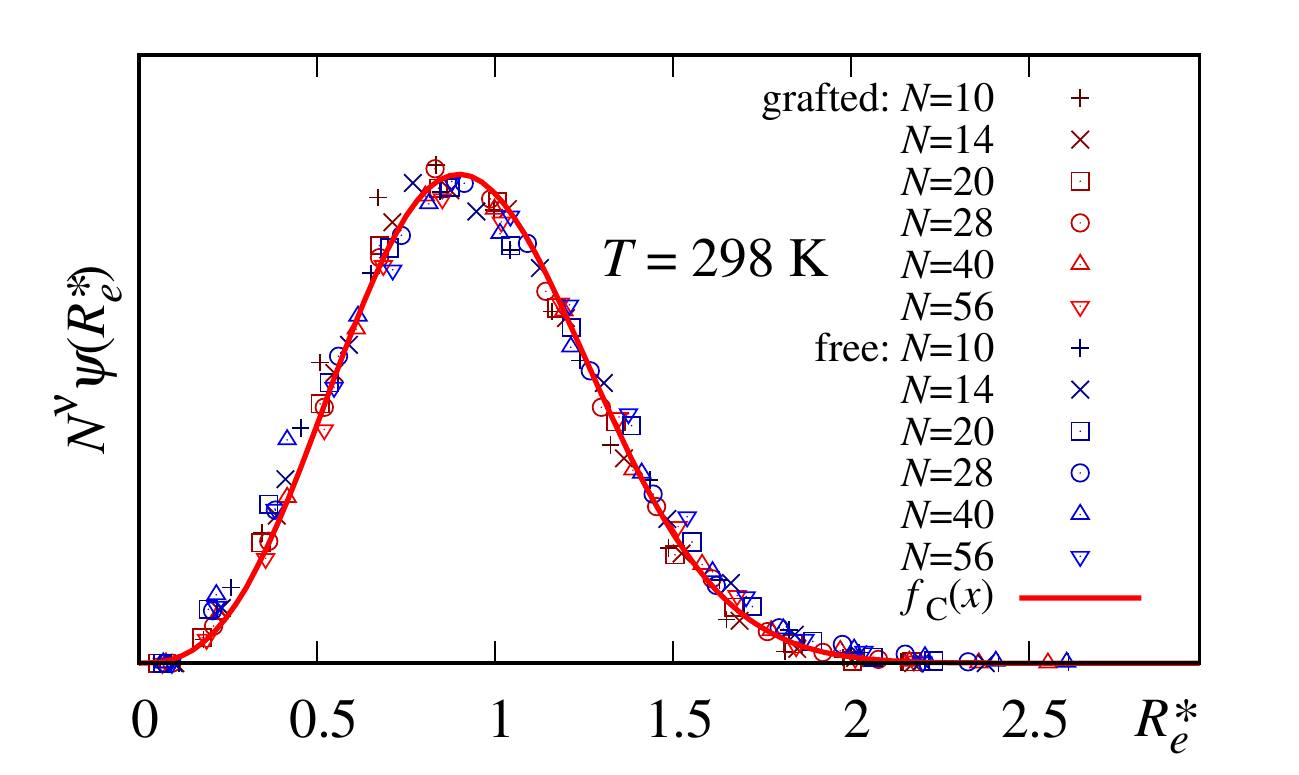}
\caption{\label{fig:Re_distr}(Colour online) Scaled distribution $N^{\nu}\psi(R^*_e)$ at $T=298\,\mathrm{K}$. Here, $R^*_e=R_e/\bar{R}_e$, where the scaling laws for $\bar{R}_e$ are given in the right-hand frame of figure~\ref{fig:Re_scaling_single}.}
\end{figure} 
Now, we discuss similar distributions for the end-to-end distance, $R_e$. Here, only the case of a good solvent, $T=298\,\mathrm{K}$, is considered. The dimensionless end-to-end distance,  $R^*_e=R_e/\bar{R}_e$, is introduced, where the scaling laws for $\bar{R}_e$ are shown in the left-hand frame of figure~\ref{fig:Re_scaling_single}. Similarly to the case of the $R_g$, scaling of the distributions $N^{\nu}\psi(R^*_e)$, obtained at different chain length $N$, enables us to introduce the master curve
\begin{equation}\label{eq:Cloiseaux}
N^{\nu}\psi(R^*_e)=f_C(R^*_e)= A_C [R^*_e]^{\delta}\exp\left(-[s_CR^*_e]^{\frac{1}{1-\nu}} \right),
\end{equation}
see figure~\ref{fig:Re_distr}. The analytic form for the function $f_C(x)$ is constructed following the analysis by des Cloizeaux~\cite{Cloizeaux1974} and de Gennes~\cite{deGennes1979}. Using numeric fits for the simulation data, we found that $A_C\approx 5.65$, $\delta\approx 2.63$ and $s_C\approx 1.15$. Since $\delta>1$, the shape of $N^{\nu}\psi(R^*_e)$ at small $R^*_e$ is convex. One should mention that the theoretical analysis~\cite{Cloizeaux1974,deGennes1979} yields the value of $\delta=(1-\gamma)/\nu\approx 0.27<1$ resulting in a concave curve at small $R^*_e$. We believe that this discrepancy is the result of the excluded volume effects at small $R_e$ in the DPD modelling approach, and these reduce the probability for the values $R^*_e\to 0$.

\begin{figure}[htb]
  \centering
\hspace{-2em}
\includegraphics[width=0.5\linewidth]{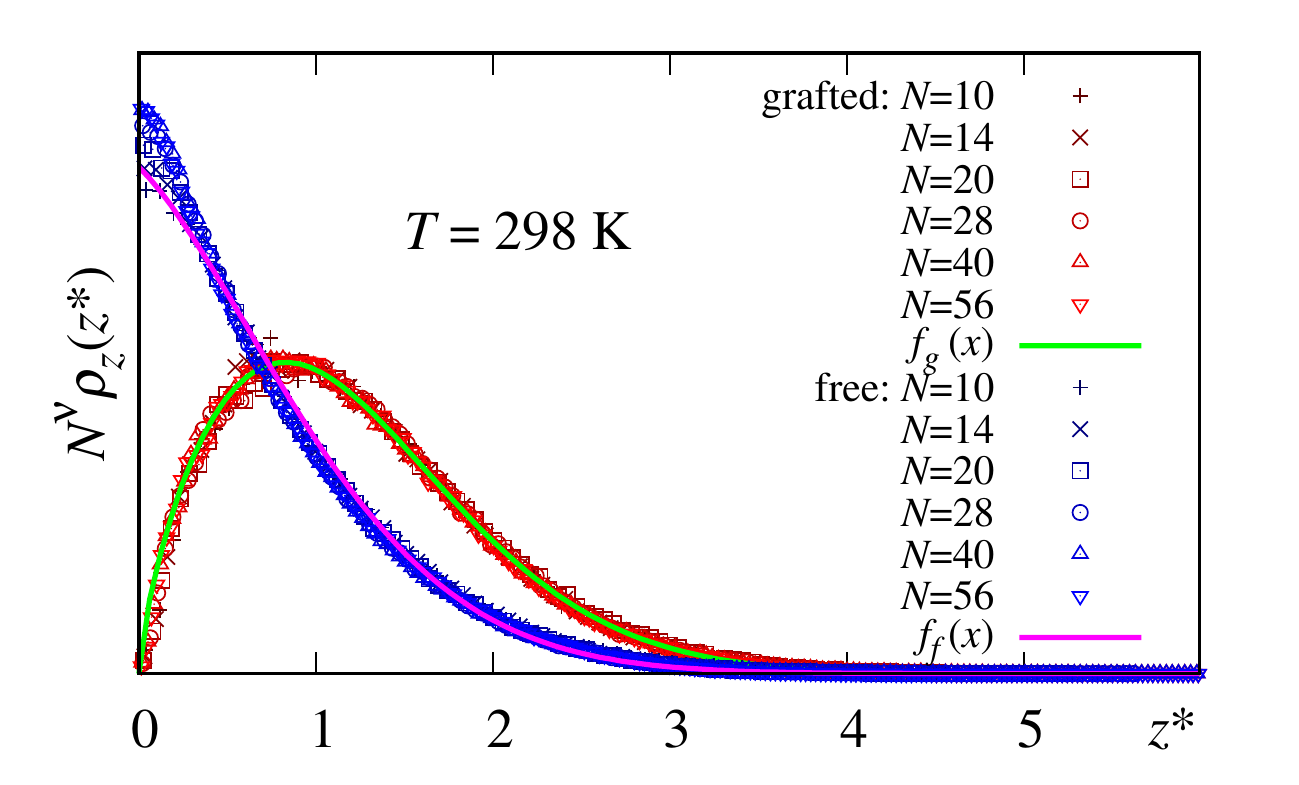}\hspace{-2em}
\includegraphics[width=0.5\linewidth]{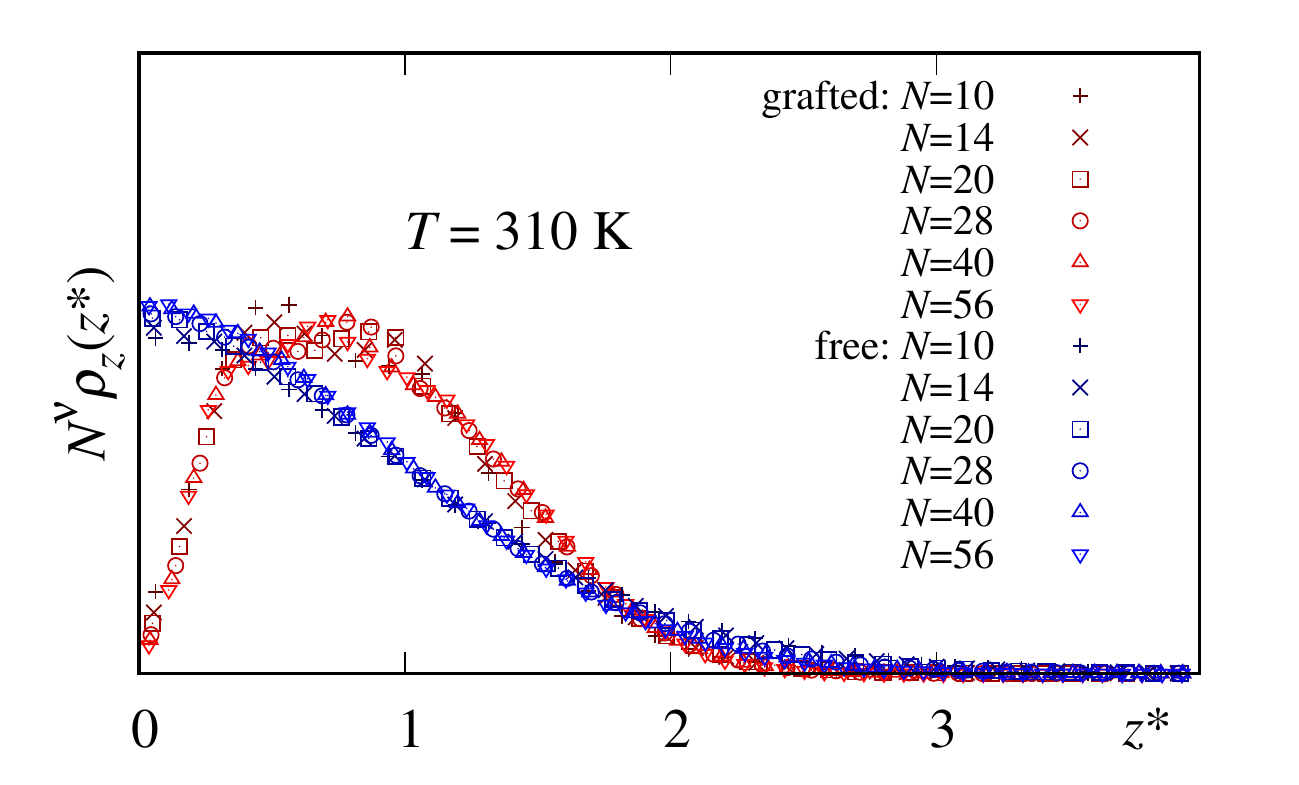}
\caption{\label{fig:zprof}(Colour online) Left-hand frame: scaled density profiles $N^{\nu}\rho_z(z^*)$ for grafted and free chain, respectively at $T=298\,\mathrm{K}$. Here, $z^*=z/\bar{R}_g$, where the scaling law for $\bar{R}_g$ follows that of figure~\ref{fig:Rg_scaling_single}. Right-hand frame: the same at $T=310\,\mathrm{K}$. Symbols represent DPD simulation data, solid lines, marked $f_g(x)$ and $f_f(x)$, are the fits to the analytic forms (\ref{eq:zprofile_srf}) and (\ref{eq:zprofile_blk}).}
\end{figure} 
Finally, we examine the density profile $\rho_z(z)$ for a single model PNIPAM chain grafted to the plane $z=0$, and compare it to its counterpart for a free chain. In the latter case, the reference plane is oriented randomly in space. It intersects the first bead of a chain and is completely transparent. For the case of $T=298\,\mathrm{K}$, the density profile can be built in terms of the dimensionless separation from the wall, $z^*=z/\bar{R}_g$ using the same scaling law for $\bar{R}_g$ as above. The data for a scaled profile, $N^{\nu}\rho_z(z^*)$, obtained at different $N$, is found to very well follow their respective master curves (for the grafted and free chain, respectively), see the left-hand frame of figure~\ref{fig:zprof}.

In the case of a grafted chain, the plane $z=0$ is impenetrable and it causes depletion of the polymer beads in its vicinity, see a respective plot in the left-hand frame of figure~\ref{fig:zprof}. The analytic form for the scaled profile $N^{\nu}\rho_z(z^*)$ can be constructed following considerations by de Gennes~\cite{deGennes1980}. In particular, at small $z$, one obtains $N^{\nu}\rho_z(z^*)\sim (z^*)^{\frac{1-\nu}{\nu}}$. At large $z^*$, we assume an exponential decay of $\rho_z(z^*)$ similarly to equation~(\ref{eq:Lhuillier}), but with unknown parameters. Then, we assume that both terms can be multiplied
\begin{equation}\label{eq:zprofile_srf}
N^{\nu}\rho_z(z^*)=f_g(z^*)=A_g (z^*)^{\frac{1-\nu}{\nu}} \exp{\left[-f_g\, (z^*)^{\alpha_g}\right]}.
\end{equation}
Numeric fits of simulation data yield $A_g\approx 0.3$, $f_g\approx 0.56$, and $\alpha_g\approx 1.74$ and the obtained master curve represents simulation data very well, see respective plot in the left-hand frame of figure~\ref{fig:zprof}.

In the case of a free chain, there is no effect from the impenetrable wall. Therefore, a profile has the Gauss-like shape, see a respective plot in the left-hand frame of figure~\ref{fig:zprof}. This leads to a guess that the analytic form for the scaled profile $\rho_z(z^*)$ in this case can be constructed by modifying the equation~(\ref {eq:zprofile_srf}) by: (i) removing the term ${z^*}^{\frac{1-\nu}{\nu}}$, and, (ii) shifting the argument in the exponential term
\begin{equation}\label{eq:zprofile_blk}
N^{\nu}\rho_z(z^*)=f_f(z^*)=A_g \exp{\left[-f_g (z^*+s_f)^{\alpha_f}\right]}.
\end{equation}
Note, that the same coefficients, $A_g$, $f_g$, and the exponent $\alpha_g$ enter this expression, whereas the amount of shift is given by $s_f\approx 0.247$. This analytic form very well describes the scaled profile $\rho_z(z^*)$ for a free chain, too, see the left-hand frame of figure~\ref{fig:zprof}.

The situation is quite similar at $T=310\,\mathrm{K}$, in which case the scaling law for $\bar{R}_g$, shown in the right-hand frame of figure~\ref{fig:Rg_scaling_single}, is used. The DPD simulation data for both grafted and free chain, represented in terms of $z^*=z/\bar{R}_g$,  reasonably  well follow their respective universal shapes $\rho_z(z^*)$, see right-hand frame of figure~\ref{fig:zprof}. One should remark that both shapes are very similar to their counterparts at $T=298\,\mathrm{K}$, see left-hand frame of the same figure, but their real separation from the wall is different  because of a different scaling exponent for $\bar{R}_g$. Similarly to the case of figure~\ref{fig:Rg_distr}, we have not performed analytic fits for the $\rho_z(z^*)$ distributions at $T=310\,\mathrm{K}$.

To conclude this section, here we performed validation of the scaling behaviour of a single chain representing the PNIPAM polymer at a mesoscopic level. Following reference~\cite{SotoFigueroa2012}, the temperature effect is modelled via the change of the repulsive strength $a_{pw}$ between PNIPAM monomers and the beads representing water, see equation~\ref{FC}. Both cases of $T=298\,\mathrm{K}$ (below the LCST), and $T=310\,\mathrm{K}$ (above it) are considered. At the first stage of this analysis, the scaling laws for the average radius of gyration, $\bar{R}_g$, and these for the end-to-end distance, $\bar{R}_e$, are obtained by performing simulations at a range of chain lengths, $N=10$--56. The results are shown in figures~\ref{fig:Rg_scaling_single} and \ref{fig:Re_scaling_single} and these indicate a reasonably good agreement with available theoretical and computational results. At the following stage, we examined the scaling properties of the distributions obtained for the $R_g$ and $R_e$ values and the density profile of monomers normally to the wall. Where known, the asymptotic scaling behaviour was used from the available literature, otherwise, suitable analytic forms were constructed leading to a very good mapping of the obtained simulation data onto master curves. Therefore, the mesoscopic model for the PNIPAM polymer employed in this study, rather accurately demonstrates the scaling properties for the size properties, $\bar{R}_g$ and $\bar{R}_e$, their distributions, and the density profile of monomers in relation to the wall.

\section{Thermoresponsive polymer brush}\label{sec4}

We proceed now to the case of a polymer brush formed of $M$ grafted chains according to the procedure explained in detail in section~\ref{sec2}. The length of all chains is fixed at $N=60$. The new parameter that is brought into a play here is the grafting number density of chains, $\rho_g$, introduced in section~\ref{sec2}. At low $\rho_g$, polymer chains behave independently of one another exhibiting the so-called ``mushroom'' regime. Here, the behaviour of chains is principally governed  by the quality of the solvent, the situation being similar to the case of an isolated chain discussed in section~\ref{sec3}. At a very high $\rho_g$, the chains are closely packed, and their properties are affected mostly by the effects of the excluded volume from the overlaps with the neighbouring chain. The crossover between these two regimes is expected at the intermediate values of~$\rho_g$. Scaling properties of polymer chains are widely covered in a number of studies~\cite{deGennes1980,Milner1988,Lai1991,Lai1992,Binder2002,Descas2008,Li2010,Binder2012}. Therefore, here we restrict our analysis to the aspects most relevant to the thermoswitching behaviour of the model chains representing the PNIPAM polymer brush.

Let us introduce some relevant properties of a brush. The effective brush height can be defined as~\cite{Binder2002}
\begin{equation}\label{eq:br_height}
\bar{h} = 2\frac{\int_{0}^{L_z}z\rho_z(z)\rd z}{\int_{0}^{L_z}\rho_z(z)\rd z},
\end{equation}
where $\rho_z(z)$ is the density profile for the brush beads, and $L_z$ is the simulation box dimension along the OZ~axis. Besides this collective feature of the ensemble of chains, we also consider individual chain characteristics related to their conformations. The size characteristics such as $\bar{R}_g$ and $\bar{R}_e$, are already introduced in equations~(\ref{eq:Rg_scaling}) and (\ref{eq:Re_scaling}). We also complement our analysis by the shape characteristics such as: the average asphericity, $A$, and the shape descriptor, $B$,~\cite{Kalyuzhnyi2016,Haydukivska2021,Haydukivska2022} defined as
\begin{equation}\label{eq:A_B}
\bar{A} = \frac{1}{6}\sum_{i=1}^{N}\left\langle \frac{\sum_{\alpha=1}^{3}(\lambda_{\alpha}-\bar{\lambda})^2}
{\bar{\lambda}^2}\right\rangle,\qquad
\bar{B} = 3\sum_{i=1}^{N} \left\langle \frac{\bar{\lambda}-\lambda_{2}}{\lambda_{1}}\right\rangle,
\end{equation}
where $\bar{\lambda}=\frac{1}{3}\sum_{\alpha=1}^{3}\lambda_{\alpha}$, and the averaging in (\ref{eq:A_B}) is performed over the time trajectory and over all chains in the brush. The benefit from the use of the shape descriptor is that it is symmetric, namely $B=-1$ for an infinitely thin disc, $B=1$ for infinitely thin rod and $B=0$ for the case of a sphere.

It is well known that with an increase of the grafting density a brush enters the so-called ``dense brush regime'', where  individual chains are stretched out and the brush height increases~\cite{Binder2012}. To quantify the difference when and to what extent this takes place at   $T=298\,\textrm{K}$ and $310\,\textrm{K}$, we consider the set of ratios
\begin{equation}
R_g^{\dagger}=\frac{\bar{R}_g(T=298\,\textrm{K})}{\bar{R}_g(T=310\,\textrm{K})},~~~
R_e^{\dagger}=\frac{\bar{R}_e(T=298\,\textrm{K})}{\bar{R}_e(T=310\,\textrm{K})},~~~
A^{\dagger}=\frac{\bar{A}(T=298\,\textrm{K})}{\bar{A}(T=310\,\textrm{K})},~~~
B^{\dagger}=\frac{\bar{B}(T=298\,\textrm{K})}{\bar{B}(T=310\,\textrm{K})},
\end{equation}
between respective values obtained at these two temperatures.

\begin{figure}[htb]
\centering
\hspace{-2em}
\includegraphics[width=0.5\linewidth]{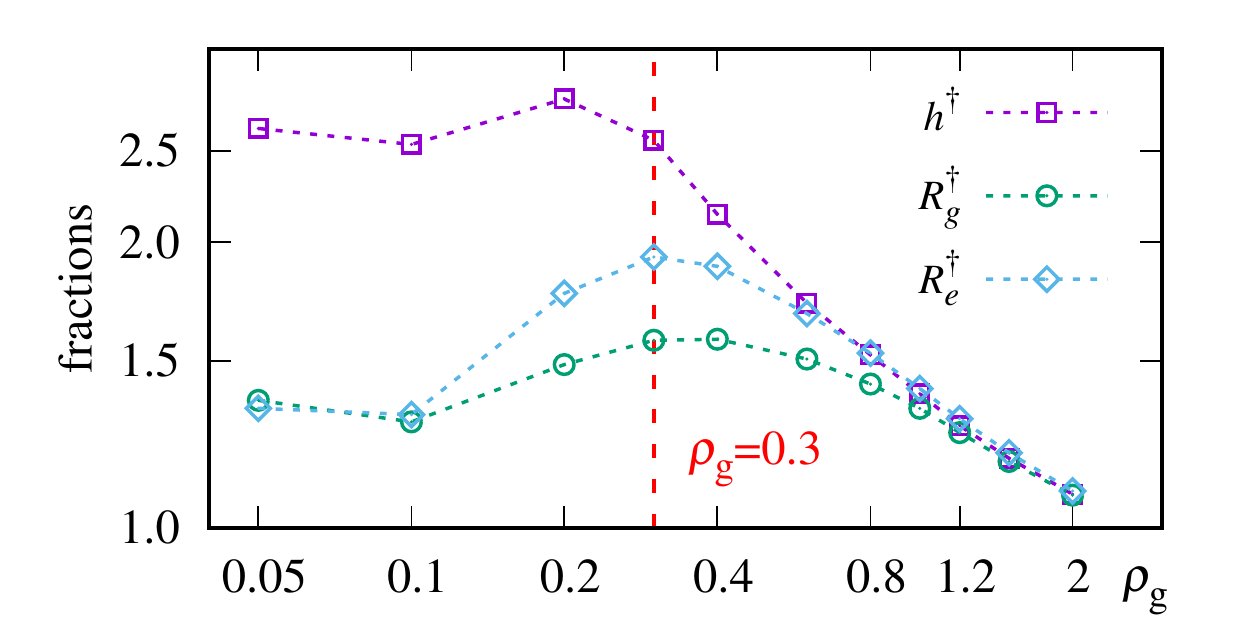}\hspace{-2em}
\includegraphics[width=0.5\linewidth]{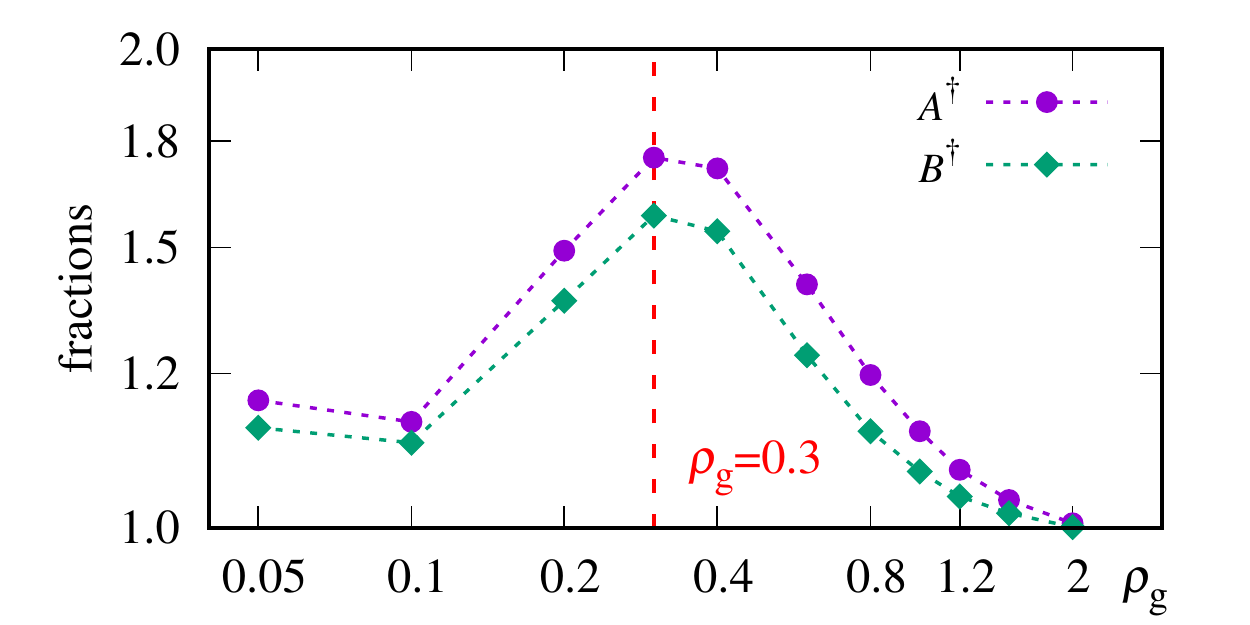}
\caption{\label{fig:h_Rg_Re_A_B_frac}(Colour online) Characteristic ratios, $h^{\dagger}$, $R_g^{\dagger}$, $R_e^{\dagger}$, $A^{\dagger}$, and $B^{\dagger}$ as the functions of the grafting density $\rho_g$. Suggested optimal grafting density, $\rho_g=0.3$, is marked by red.}
\end{figure} 
The dependencies of these ratios on the grafting density are shown in  figure~\ref{fig:h_Rg_Re_A_B_frac}. The effectiveness of the thermoresponsive brush as a functional surface can be measured via the magnitude of the $h^{\dagger}$ ratio, which is found to be the largest at $\rho_g\leqslant 0.3$. One can also see that all the other fractions, $R_g^{\dagger}$, $R_e^{\dagger}$, $A^{\dagger}$, and $B^{\dagger}$, reach their maximum values at $\rho_g\approx 0.3$. This hints at this interval of grafting densities as the one where the difference between two temperatures in terms of the amount of a stretch of individual chains is the largest.

\begin{figure}[htb]
  \centering
\includegraphics[width=0.45\linewidth]{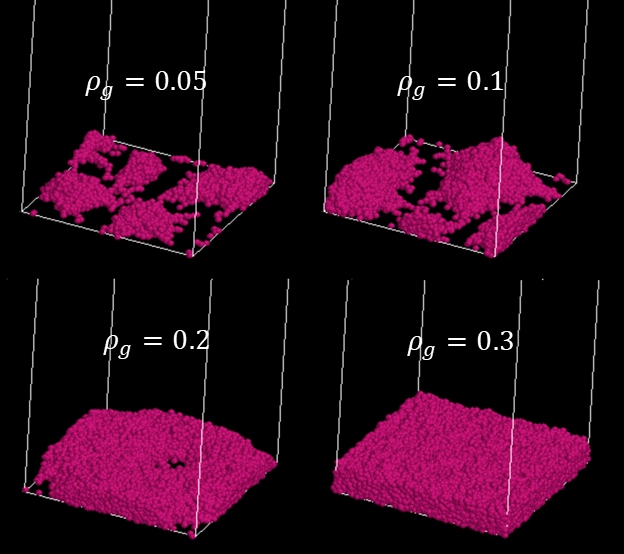}
\caption{\label{fig:brush_snapshots}(Colour online) Snapshots of the polymer brush structure (shown in pink) for low and medium densities, $\rho_g=0.05$--0.3, at $T=310\,\textrm{K}$. Note inhomogeneous structure of a brush for $\rho_g=0.05$--0.2.}
\end{figure} 
The narrow interval of grafting densities around $\rho_g\approx 0.3$ is a good candidate for designing a thermoswitchable brush from an additional reason as well. Figure~\ref{fig:brush_snapshots} shows the structure of a brush surface at $T=310\,\textrm{K}$, in the collapsed state. It can be seen that the surface is very inhomogeneous showing islands of collapsed chains and the regions of exposed surface until the grafting density reaches the value of $\rho_g\approx 0.3$. Such a behaviour at low grafting densities is well known since the early simulation studies~\cite{Lai1992,Grest1993}. In most practical applications, such inhomogeneity would be rather unwelcome as it may hamper the adsorption of nanoparticles in the collapsed state of a brush. Additionally, application of the PNIPAM-based brush as a part of a mixed brush system for thermocontrolled adsorption/desorption of proteins~\cite{Kim2021} requires the highest possible grafting density of the PNIPAM component to ensure strong push-effect at the desorption stage. These considerations lead us to believe that the grafting densities close to the $\rho_g\approx 0.3$ are optimum for the application of a thermoresponsive brush as an effective height-changing functional surface.

Let us examine the origins of the differences in the brush behaviour at various grafting densities. To this end, the blob interpretation of polymer chains by Alexander and de Gennes~\cite{Alexander1977,deGennes1980} can be invoked. In brief, if the average separation between the grafted points, $d$, is getting smaller than the average radius of gyration of an isolated chain, $\bar{R}_g$, then the chain monomers started to redistribute their mass normally to the surface because of the excluded volume effect. The chain can be interpreted as a linear stack of blobs of diameter $d$, assembled normally to the surface, where the number of monomers $n$ in one blob can be estimated from the relation $d=l\,n^{\nu}$, where $\nu=0.5882$, $0.5$, and $1/3$ for the good solvent, $\theta$-point and poor solvent regime, respectively, inside each blob. The blob interpretation leads to the following scaling law for $\bar{h}$~\cite{deGennes1980}
\begin{equation}\label{eq:brush_scaling}
\bar{h} \sim c\,\rho_g^\epsilon\,N, \qquad \epsilon={\frac{1-\nu}{2\nu}},
\end{equation}
where $\epsilon\approx 0.35$ for the case of a good solvent within a blob, and $\epsilon=1/2$ for the case of a $\theta$-point within a blob (desolvated chain). Similar scaling laws are also expected for both the values of $\bar{R}_g$ and $\bar{R}_e$ for each chain in a brush. 

\begin{figure}[htb]
\centering
\hspace{-2em}
\includegraphics[width=0.5\linewidth]{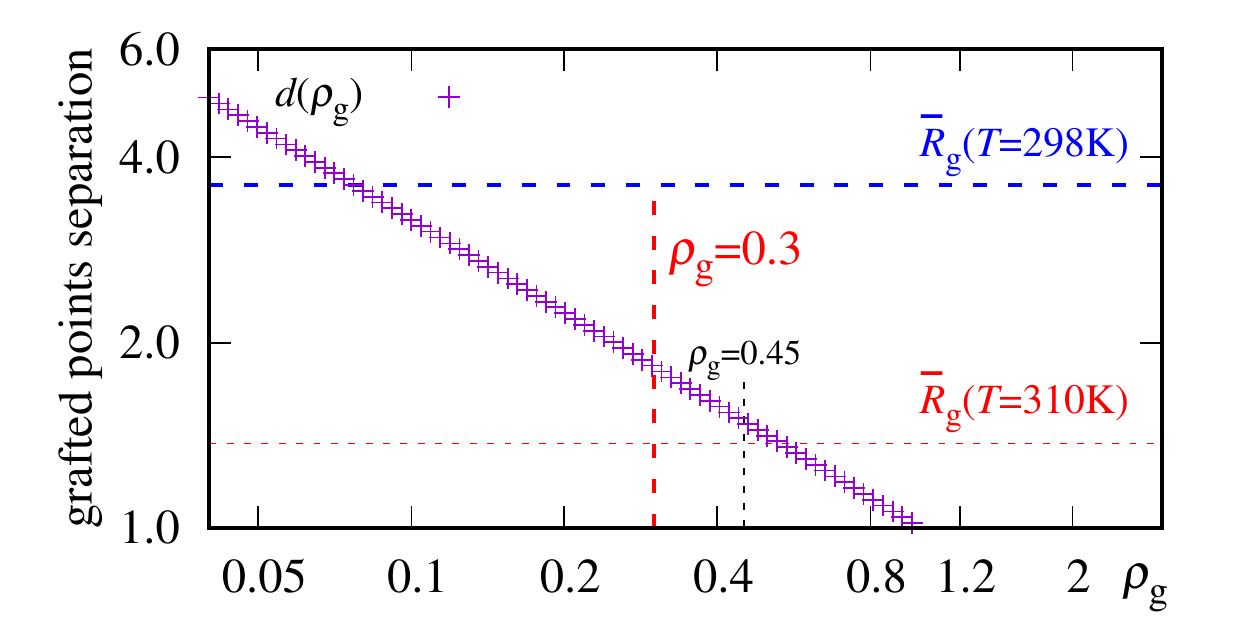}\hspace{-2em}
\includegraphics[width=0.5\linewidth]{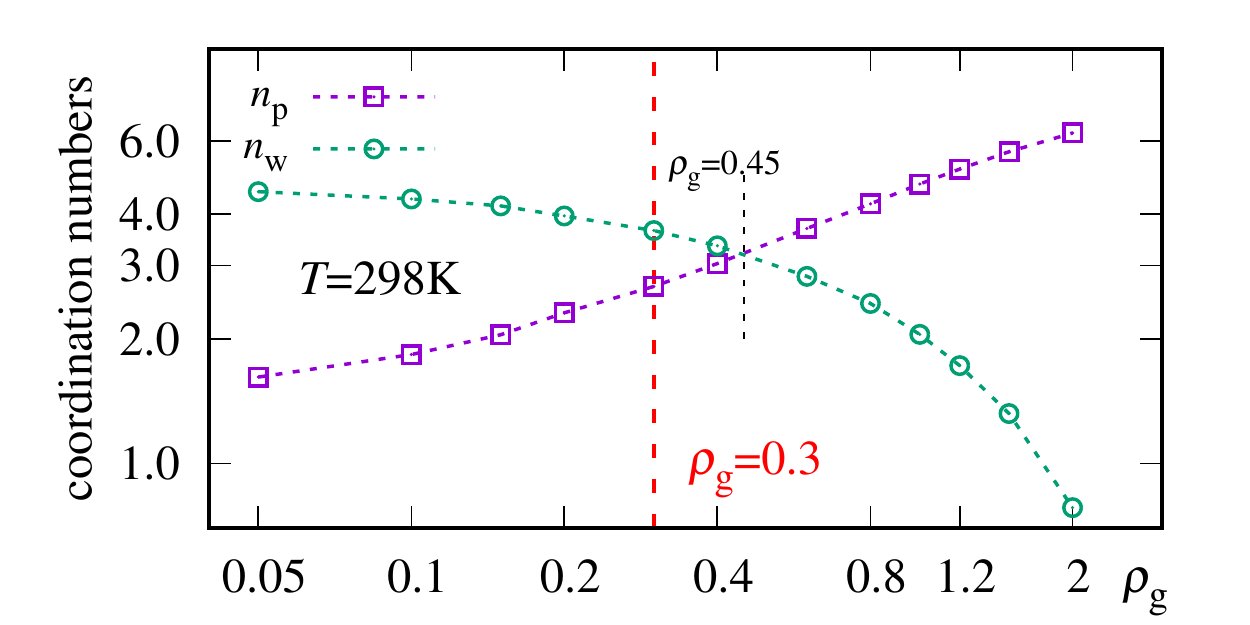}
\caption{\label{fig:solvent}(Colour online) Left-hand frame: dependence of the separation $d$ between grafted points of a brush on grafting density $\rho_g$. The values for the radius of gyration of isolated chain, $\bar{R}_g$, at $T=298\,\textrm{K}$ and $310\,\textrm{K}$, based on figure~\ref{fig:Rg_scaling_single}, are shown as horizontal dashed lines. Right-hand frame: the same for the partial coordination numbers, $n_p$ and $n_w$, for polymer and water in the vicinity of each monomer of a chain ($T=298\,\textrm{K}$).}
\end{figure} 
Hence, the changes in the brush height upon increase of $\rho_g$ are governed by two factors. The first one is the excluded volume effect between the neighbouring chains, that is brought into play when $d(\rho_g)$ decreases below the $R_g$ value. The dependence of $d(\rho_g)$ on $\rho_g$ is easily established, $d(\rho_g)=\rho_g^{-1/2}$, and it is shown in the left-hand frame of figure~\ref{fig:solvent}. As one can see, at $T=298\,\textrm{K}$ the value of $d$ drops below $R_g$ at $\rho_g\approx 0.07$, whereas at $T=310\,\textrm{K}$ this occurs for a much higher grafting density, $\rho_g\approx 0.5$. As the result, at grafting density 
$\rho_g=0.3$, the brush at $T=298\,\mathrm{K}$ is already deep into the 
dense brush regime, whereas at $T=310\,\mathrm{K}$ it is still in the 
``mushroom'' state, thus explaining a large difference in its height at 
these respective temperatures.

The second factor is the solvation regime of chain monomers within each blob, which affects the value of the exponent $\epsilon$. To clarify this factor, we performed analysis of the contents of the first coordination sphere for all monomers at each grafting density. The radius of the first coordination sphere was chosen equal to the cutoff distance of the conservative force (\ref{FC}), and the number of polymeric monomers and that for solvent beads were counted. These were subsequently averaged over all existing monomers yielding $n_p$ and $n_w$, the partial coordination numbers associated with the polymer-polymer and polymer-solvent pairs. Their dependencies on the grafting density $\rho_g$ are shown in the right-hand frame of figure~\ref{fig:solvent}. Total number of beads in the first coordination sphere, $n_p+n_w$, is close to six, as expected. At a low grafting density, one has strong solvation of polymer beads, $n_s \gg n_p$, whereas strong desolvation occurs at high grafting densities, where $n_s \ll n_p$. Crossover takes place at the grafting densities close to $\rho_g\approx 0.45$ (see, black dashed line in figure~\ref{fig:solvent}). Let us note that at $\rho_g=0.3$ (i.e., the supposed optimum grafting density for the thermoswitchable brush) the polymer chains are still, at least partially, solvated.

\begin{figure}[htb]
  \centering
\includegraphics[width=0.6\linewidth]{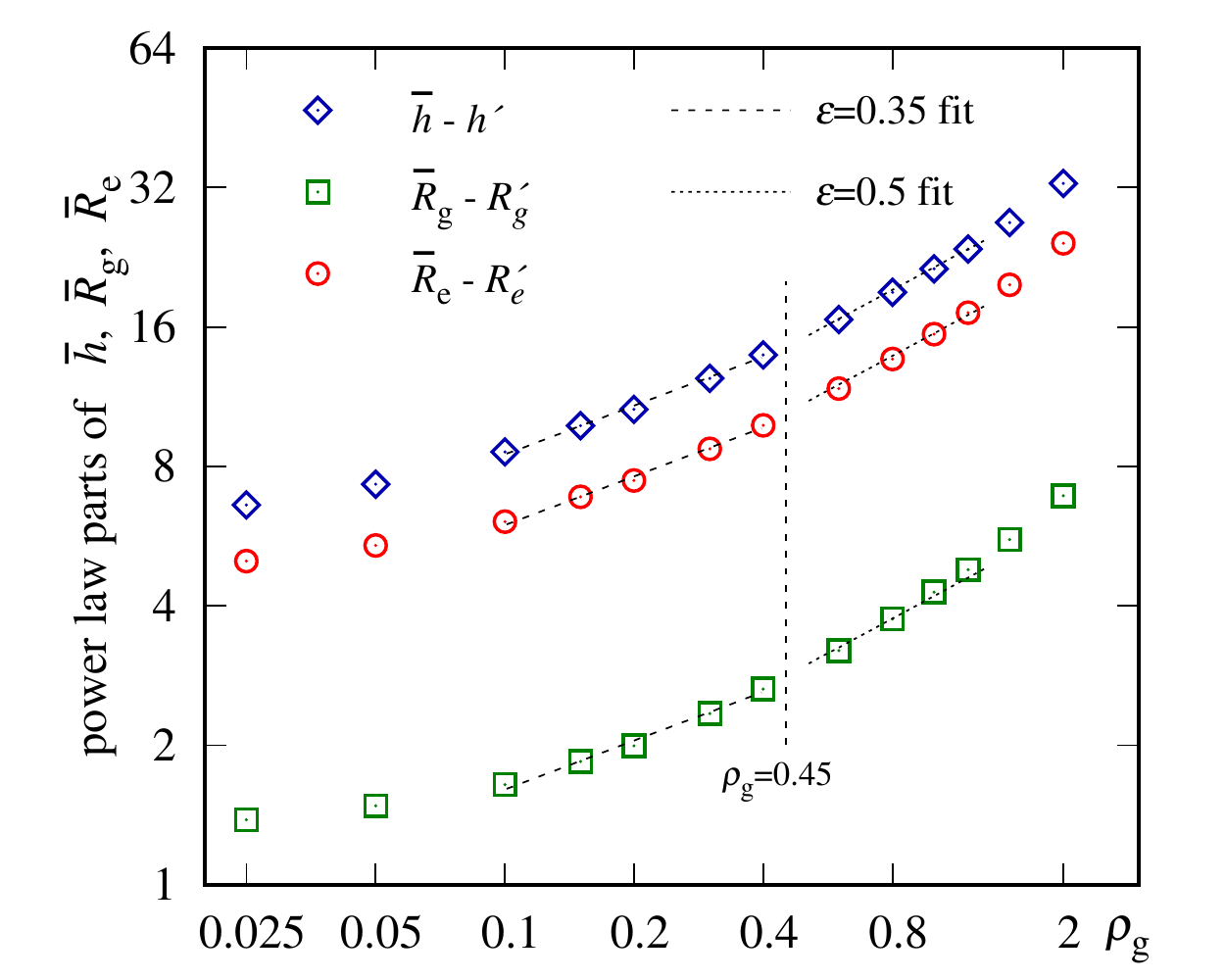}
\caption{\label{fig:brush_scaling}(Colour online) Terms proportional to $\rho_g^{\epsilon}$ in the expressions for $\bar{h}$, $\bar{R}_g$, and $\bar{R}_e$, see equation~(\ref{eq:brush_scaling_2}), and their fits at fixed values of the exponent $\epsilon=0.35$ and $0.5$.}
\end{figure} 
Now, after the solvation regimes at various grafting densities are clarified, the scaling laws of the form (\ref{eq:brush_scaling}) can be verified. To do so, we take into account that $N=60=\textrm{const}$ in current simulations, and assume the presence of some constant terms, $h^0$, $R_g^0$, and $R_h^0$, in the scaling expressions 
\begin{equation}\label{eq:brush_scaling_2}
\bar{h} = h^0 + C_h\,\rho_g^\epsilon, \qquad \bar{R}_g = R_g^0 + C_g\,\rho_g^\epsilon, \qquad \bar{R}_e = R_e^0 + C_e\,\rho_g^\epsilon.
\end{equation}
Then, we attempted to find the respective intervals for $\rho_g$, where the fits with $\epsilon=0.35$ (solvation regime) and $\epsilon=0.5$ (desolvation regime) are reasonably accurate. The results are shown in figure~\ref{fig:brush_scaling}. The following values for the coefficients entering equation~(\ref{eq:brush_scaling_2}) are found: $h^0\approx 2.9$, $C_h\approx 19.0$, $R_g^0\approx 2.2$, $C_g\approx 3.6$, $R_e^0\approx 4.2$, and $C_e\approx 13.4$. The intervals being found are: $0.1<\rho_g<0.5$ for the $\epsilon=0.35$ fit and $0.5<\rho_g<1.3$ for the $\epsilon=0.5$ fit. These are in a good agreement with the picture of solvation regimes, as identified in figure~\ref{fig:solvent}. Therefore, we may conclude that, in general, the model polymer brush discussed here follow the scaling laws for the brush height, radius of gyration and the end-to-end distance, within the blob interpretation suggested by Alexander and de Gennes~\cite{Alexander1977,deGennes1980,Descas2008,Binder2012}.

Therefore, computer simulations of the model thermoresponsive brush considered in this study reveals that there is some optimum grafting density, $\rho_g\approx 0.3$, at which the ratio between the brush height at $T=298\,\textrm{K}$ and $310\,\textrm{K}$ is close to its maximum value (see, figure~\ref{fig:h_Rg_Re_A_B_frac}) and the structure of the brush in its collapsed state is homogeneous (see, figure~\ref{fig:brush_snapshots}). There is a combination of two factors that contribute to such a difference between the state of the brush for $\rho_g=0.3$ and at these two temperatures. The first one is related to the excluded volume effect between the neighbouring chains for $\rho_g=0.3$. It already promotes the increase of the brush height at $T=298\,\textrm{K}$ [$d(\rho_g)<\bar{R}_g$], but has no influence on it at $310\,\textrm{K}$ [$d(\rho_g)>\bar{R}_g$], see left-hand frame of figure~\ref{fig:solvent}. The second factor is related to the fact that the monomers are, at least partially, solvated for $\rho_g=0.3$ at $T=298\,\textrm{K}$, see right frame of figure~\ref{fig:solvent}. This causes the chains to swell more as compared to the desolvated chains at $T=310\,\textrm{K}$. With the increase of $\rho_g$, the chains at $T=298\,\textrm{K}$ became desolvated as well, thus decreasing the difference between the cases of $T=298\,\textrm{K}$ and $310\,\textrm{K}$. With an increase of the grafting density towards dense brush regime, brush chains became equally desolvated and stretched both at $T=298\,\textrm{K}$ (below LCST) and $T=310\,\textrm{K}$ (above LCST), and the difference between these two temperatures vanishes, see, figure~\ref{fig:h_Rg_Re_A_B_frac}.

\section{Conclusions}\label{sec5}

Thermoresponsive polymer brushes, based on the PNIPAM polymer, are widely used in a number of practical applications. These include drug delivery systems, protein and microalgae adsorption/desorption, membrane based filtering, etc. Their funtionality is based on the change of hydrophilicity of PNIPAM chains in water solution when the temperature crosses the LCST point. In particular, below LCST, PNIPAM is hydrophilic due to its capability to form hydrogen bonds with water molecules and, as a result, its chains adopt swollen coil-like conformation. Above the LCST, PNIPAM turns into a hydrophobic polymer and its chains collapse into a globule. Clear understanding of all the details of such coil-to-globule transition, covering the role of the chain length, details of molecular architecture, brush density and other factors, are of much benefit for designing new thermoresponsive functional materials.

Computer simulations are capable of shedding much light towards clarifying these aspects, but modelling the functionality of the thermoresponsive brush faces a number of challenges related to the need to cover relatively large time- and length-scales. In this study we employed the mesoscale simulation approach of the DPD with the use of the parametrisation for the PNIPAM chain that was performed earlier~\cite{SotoFigueroa2012}. In this approach, the temperature dependent hydrophilicity is mapped onto the effective parameters for the polymer-solvent interaction. The principal aims of our study were to test the simple linear model for the PNIPAM polymer both below and above the LCST, and to get some insight towards the optimal brush structure to ensure its functioning as a thermoswitchable functional surface. 

For the case of isolated grafted chain, we examined its scaling properties above and below the LCST and found the scaling laws to  very well match the known results. In particular, the model reproduces the correct scaling behaviour for the radius of gyration, end-to-end distance, as well as distributions of their values and the density profile with respect to the distance to the wall.

For the case of a brush, the model reveals the presence of an optimum grafting density, $\rho_g\approx 0.3$, at which the ratio between the brush height below LCST and above it reaches its maximum value, indicating the largest amplitude of a thermoswitching effect. We note that this ratio is about $2.5$, close to some reported experimental values~\cite{Kim2021}. At this grafting density, similar ratios, evaluated for the radius of gyration, end-to-end distance, asphericity, and the shape factor of individual chains, all reach their respective maxima. The reason for such a behaviour is examined by using the concept of blobs for a polymer brush by Alexander and de Gennes~\cite{Alexander1977,deGennes1980}. At a low grafting density, the brush is in a mushroom regime, i.e., it comprises a set of non-interacting chains. At a higher grafting density, the excluded volume effect between the neighbouring chains became important and causes the stretching of the polymer chain away from a wall leading to the interpretation of each chain as a linear stack of blobs. We found that at the grafting densities of $\rho_g\approx 0.3$, there is an essential difference between the chain conformations below and above the LCST. In particular, below LCST, the chains of a brush are deep in a stretched regime, whereas above LCST, the chains are still in a mushroom regime. This difference in chain conformations is further amplified by a good solvent condition below LCST and by poor solvent conditions above it. Both factors combined are responsible for the largest amplitude of a thermoswitching effect observed at this grafting density.

The results of this study can be used in a number of ways. First, they open up the possibility to model of the temperature controlled adsorption/desorption of proteins, as, e.g., in the experimental work reported in reference~\cite{Kim2021}, or other macromolecules. The results can also be used to predict an optimal grafting density of a thermoresponsive polymer brush. For instance, one may perform the measurements of the radius of gyration of isolated chains first, and then use these to estimate the grafting density when the mushroom regime switches into the stretched one. Other applications of the  presented results are possible as well.

\section*{Acknowledgements}
DY and JI thank the CRDF for financial support under Award number G-202203-68625. DY and OK acknowledge partial support from the NAS of Ukraine (the grant for research laboratories/groups of young scientists No 07/01-2022(4)). JI thanks to S.~Minko, T.~Patsahan, and O.~Zaichenko for useful discussions. All authors are grateful to the Armed Forces of Ukraine for the protection during this research work.


\newpage

\ukrainianpart
\title{Моделювання термочутливої полімерної щітки за допомогою мезоскопічного методу комп’ютерного моделювання}
\author{Д. Яремчук, О. Калюжний, Я. Ільницький}
\address{Iнститут фiзики конденсованих систем Нацiональної академiї наук України, вул. Свєнцiцького, 1, 79011, Львiв, Україна }
\makeukrtitle
\begin{abstract}
Ми розглядаємо функціональну поверхню, що містить термочутливі полімерні ланцюги, матеріал, який широко застосовується в технології та біомедицині. З метою опису її характерних просторово-часових масштабів, комп’ютерне моделювання виконується із залученням огрублених мезоскопічних моделей. Використана тут модель була запропонована авторами роботи [Soto-Figueroa та ін., Soft Matter, \textbf{8}, 1871 (2012)], і вона описує основну особливість полі(N-ізопропілакриламіду) (PNIPAM), а саме -- швидку зміну його гідрофільності при переході через нижню критичну температуру розчинності (LCST). Для випадку ізольованого ланцюга ми досліджуємо властивості масштабування його радіусу гірації, відстані між першим та останнім мономерами, низку функцій розподілу вказаних характеристик та профілі густини мономерів залежно від відстані від поверхні як нижче, так і вище за LCST. Для випадку моделі термочутливої полімерної щітки дослідження концентруються на пошуку оптимальної густини пришпилення, за якої зміна висоти щітки при переході через LCST досягає максимального значення. Надано інтерпретацію механіз\-му зміни конформацій ланцюжків щітки в термінах блобів Александера-де Жена та ступеня сольватації полімерних ланцюгів у щітці.
\keywords термочутливі полімери, PNIPAM, дисипативна динаміка
\end{abstract}

\end{document}